\documentclass[preprint,onecolumn,floatfix,superscriptaddress,amsmath,amsfonts,amssymb,aps]{revtex4}

\usepackage{graphicx}
\usepackage{dcolumn}
\usepackage{bm}
\usepackage[normalem]{ulem}
\usepackage{color}
\usepackage{amssymb}
\usepackage{amstext}

\begin{document}


\title{\mbox{Raman scattering study and lattice-dynamics calculations in YTiO$_3$:}\\ \mbox{Precursor of the magnetic phase transition in the phonon anomalies}}

\author{N. N. Kovaleva}
\email{kovalevann@lebedev.ru}
\affiliation{P.N. Lebedev Physical Institute of the Russian Academy of Sciences, Leninsky prospekt 53, 119991 Moscow, Russia}
\author{O. E. Kusmartseva}
\affiliation{Department of Physics, Loughborough University, LE11 3TU Leicestershire, United Kingdom}
\affiliation{Mediterranean Institute of Fundamental Physics, Via Appia Nuova 31, 00040 Marino, Rome, Italy}
\author{A. Maljuk}
\affiliation{Leibniz-Institut f\"ur Festk\"orper- und 
Werkstoffforschung, Helmholtzstr. 20, D-01171 Dresden, Germany}

\date{\today}

\begin{abstract}
The origin of the low-temperature ferromagnetic instability in YTiO$_3$ driven by the interplay between lattice, orbital, spin, and charge degrees of freedom is on the background of lattice dynamics. We present a comprehensive study of temperature-dependent polarized Raman scattering in a nearly stoichiometric orthorhombic YTiO$_3$ single crystal. The lattice-dynamics calculations authenticate the assignment for the observed $A_g$- and $B_{2g}$-symmetry modes in terms of the atoms vibrational patterns. The obtained results suggest that trigonal GdFeO$_3$-type distortions become unstable against tetragonal Jahn-Teller distortions, leading to the consequent orbital structure refinement and ferromagnetic ground state in YTiO$_3$.
\end{abstract}

\pacs{Valid PACS appear here}
\maketitle
\section{Introduction}\label{1}
Orbital ordering phenomena in correlated electron systems with electron-lattice interaction remain the most vital topic in condensed matter physics despite many years of scientific effort. The orbital ordering phenomena are quite well understood for representative magnetic compounds with Jahn-Teller ions, such as Mn$^{3+}$ ($3d^4$) or Cu$^{2+}$ ($3d^9$), which display many outstanding properties, including colossal magnetoresistance (manganites) and high temperature superconductivity (cuprates). The physics of these compounds is determined by the interplay between electron-lattice Jahn-Teller (JT) interaction 
\cite{JahnTeller,OpikPryce,Stoneham,BVO,KaplanVekhter,Bersuker,Bersuker_ChemRev,Kovaleva_Raman} and purely electronic superexchange (SE) interaction 
\cite{Goodenough,Kanamori,Khomskii,Oles,Kovaleva_lmo_prl,Kovaleva_lmo_prb}, which lead to lowering-symmetry structural and magnetic phase transitions. Here we begin with some sort of insight about these different mechanisms: (i) the electron-lattice JT mechanism is related to electronically-degenerate systems intrinsically unstable against asymmetric distortion, being on the background of the lattice with which these electronic degrees of freedom strongly interact, and (ii) the SE interactions determined by the on-site Coulomb repulsion $U$ are mediated by the intersite $d_id_j$ charge excitations along a bond $\left\langle ij \right\rangle$ between magnetic ions and involve both spin and orbital degrees of freedom. Generally, correlated electron systems with electron-lattice interaction exhibit a variety of physical properties governed by the interplay between the lattice, orbital, spin, and charge degrees of freedom.

The $3d(t_{2g}^1)$ series of orthorhombic ($Pnma$ space group) rare-earth $R$TiO$_3$ titanates represents a paradigm in which the end-point compounds (for $R$\,=\,La and Y) exhibit a progressing structural distortion and different orbital ordering physics, which, in turn, has an impact on their magnetic structure: LaTiO$_3$ is a G-type antiferromagnet (AFM) with the N\'eel temperature $T_{\rm N}$\,=\,150\,K \cite{Cwik}, while YTiO$_3$ is a ferromagnet (FM) with a low Curie temperature  of $T_{\rm C}$\,=\,30\,K \cite{Garett}. 
Recent experiments discovered that the FM transition temperature 
decreases in the isovalently substituted Y$_{1-x}$La$_x$TiO$_3$ 
on approaching the FM-AFM phase boundary at the La concentration 
$x_c$\,$\approx$\,0.3 \cite{Hameed}. In the idealized cubic perovskite structure ($Pm\bar3m$) of $R$TiO$_3$, featuring a chain of the corner-sharing TiO$_6$ octahedra, the ground state of Ti$^{3+}$(3$d^1$) ions in sixfold oxygen configuration in the $t_{2g}$ triplet is orbitally degenerate. The orbital degeneracy is essential for the electron-lattice, or vibronic, coupling JT effects. It was proposed that the trigonal ($D_{3d}$) distortion of the TiO$_6$ octahedra splits the threefold degenerate $t_{2g}$ levels into a non-degenerate lower $a_{1g}$ level and twofold degenerate higher $e_g$ levels \cite{Mochizuki}. However, this $D_{3d}$ distortion has not been supported by any evidence so far. In addition, it was suggested that the generic GdFeO$_3$-type distortion (which can be associated with rotations of the TiO$_6$ octahedra) promotes lift of the $t_{2g}$ orbital degeneracy by the crystal field of $R$ ions \cite{Mochizuki1,Pavarini}. This new mechanism generates a crystal field similar to the $D_{3d}$ crystal filed, which is supposed to split the $t_{2g}$ levels into three non-degenerate levels. In this case, the lowest orbital occupation would stabilize the $G$-type AFM state. This mechanism competes with the JT mechanism in YTiO$_3$ with a large JT distortion, where the latter may take control over the spin-orbital structure. However, clear 
theoretical evidence of the individual role of each lattice distortion, 
including the generic GdFeO$_3$-type oxygen octahedra rotations, 
is still missing.

Both experimental and theoretical studies of the lattice dynamics are capable of shedding light upon the role played by the JT effect on the underlying physics in YTiO$_3$ crystal. Yet only little evidence on the properties of the Raman-active modes in YTiO$_3$ are thus far available from the existing experimental and theoretical studies (see \cite{Sugai,Chernyshev} and references therein). Earlier, we demonstrated that the shell-model (SM) approach combined with interatomic potentials \cite{GULP,GULP1,Catlow} offers a powerful route to understanding optical and lattice-dynamics properties of many complex oxides \cite{Kovaleva_JETP,Kovaleva_Bi,Caimi}. Moreover, in the framework of the SM approach, we elaborated SM parameters yielding good agreement between the experimental and calculated properties for the lattice parameters, static and high-frequency dielectric constants, as well as the frequencies of the infrared (IR) phonon modes in orthorhombic YTiO$_3$ crystal \cite{KovalevaYTO_IRphon}.
   
In this work, we use high-quality nearly stoichiometric YTiO$_3$ single crystals to study the lattice dynamics by Raman spectroscopy in a broad temperature range. We analyze temperature dependences of the frequencies, widths, and integral intensities ratio of the Raman phonon bands. In addition, we perform theoretical lattice-dynamics SM calculations for the Raman normal modes in orthorhombic ($Pnma$) YTiO$_3$ crystal and provide their assignments with the experimentally observed frequencies in terms of the atom's vibrational patterns. The role played by the JT effect on the underlying physics in YTiO$_3$ crystal is discussed.     

\section{Samples and Methods}\label{2}
Single crystals of YTiO$_{3}$ were grown by the floating zone method in a reducing atmosphere (Ar/H$_2$=50/50). The apparatus used for the crystal growth was a four-mirror-type infrared image furnace (Crystal System Corp., FZ-T-10 000-H-III-VPR) equipped with four 1.5 kW halogen lamps. X-ray diffraction confirmed that the grown YTiO$_3$ single crystals are non-twinned, with mosaicity less than 0.03$^\circ$. Due to the Ti$^{3+}$$\rightarrow$Ti$^{4+}$ instability, single crystals of YTiO$_{3+\delta}$ possess an excess of oxygen, which is estimated at a level lower than 0.013 (for more details, see Refs.\,\cite{KovalevaYTO_IRphon,KovalevaYTO}). The grown ingots were aligned along the principal axes and cut in the form of a parallelepiped with approximate dimensions of $\sim$3$\times$3$\times$3 mm$^3$. We note that one should avoid any warming above 450$^\circ$C during cutting and polishing of the samples, since this can lead to the stabilization of Y$_2$Ti$_2$O$_7$ compound.

For optical measurements, the sample surfaces were polished to optical grade. Raman scattering spectra were measured using a Raman-microscope spectrometer LabRAM HR (Horiba Jobin Yvon) equipped with a grating monochromator and a liquid-nitrogen cooled CCD detector. A HeNe laser operating at 632.8 nm wavelength was used as the excitation light source. The crystal sample was mounted on a cold-finger of the microcryostat cooled with liquid nitrogen. The near-normal back scattering geometry was used. The propagation direction of the incident and scattered light was parallel to the chosen principal $Pnma$ orthorhombic axis. The principal polarizations c(bb)c and b(ca)b were measured using generic linear polarization of laser irradiation. The Raman scattering tensor determines whether the normal phonon modes actually appear on the oriented sample surface under irradiation with polarized laser light at a given scattering geometry. In the parallel (aa), (bb), and (cc) polarizations of the incident and scattered light, the A$_{\rm g}$ symmetry phonon modes show up, whereas in the crossed (ba,ab), (ca,ac), and (bc,cb) polarizations the B$_{\rm 1g}$, B$_{\rm 2g}$, and B$_{\rm 3g}$ symmetry modes could appear, respectively. To obtain temperature dependences of the Raman modes' resonant frequencies, their full widths at half maximum (FWHM), and the integral intensities, the polarized Raman spectra measured at different temperatures were fitted with Lorentzian bands.

To make an assignment of the phonon modes in the experimental Raman scattering spectra, we performed lattice-dynamics calculations for an orthorhombic YTiO$_3$ crystal in the $Pnma$ space group (D$^{16}_{2h}$, No.\,62) in the SM approach. The present calculations were performed in the General Utility Lattice Program (GULP) code \cite{GULP,GULP1}. For the program input we used the unit cell parameters and fractional coordinates of the atoms for YTiO$_3$ single crystals grown by the Czochralski method \cite{Maclean}. In the SM framework, the lattice is represented by polarizable ions constructed from massive point cores and massless shells, which are coupled by isotropic harmonic forces defined by spring constants. The interaction includes contributions of the Coulomb, polarization, and short-range interactions. The electronic polarization of the ions is represented by the displacement of their shells relative to the cores. In our model, cations are treated as non-polarizable and the short-range interactions between them were ignored. The short-range potentials for the shell-shell (oxygen-oxygen) and core-shell (cation-oxygen) interactions are of the Buckingham form,  
\begin{equation}
V_{ij}=A_{ij}{\rm exp}(-r/\rho_{ij})-C_{ij}/r^6. 
\end{equation}
The parameters of both repulsive and attractive components of the Buckingham potential for the O-O (shell-shell) interactions were taken as those of typical oxides \cite{Catlow}. The Buckingham parameters for the cation-shell interactions, shell charges, and force constants were fitted using the lattice parameters, the static and high-frequency dielectric constants, and the frequencies of the transverse optical (TO) phonon modes. The resulting Buckingham potential parameters ($A_{ij}$, $\rho_{ij}$, and $C_{ij}$), shell charges, and force constants are presented in Table III of our previously published article focused on the study of the IR-active phonon modes in YTiO$_3$ single crystal \cite{KovalevaYTO_IRphon}. We demonstrated that the elaborated SM parameters give good agreement between the experimental and calculated properties for the lattice parameters, static and high-frequency dielectric constants, as well as the frequencies of the TO phonon modes (see Tables IV and VI in Ref. \cite{KovalevaYTO_IRphon}). In the present study, using the elaborated SM parameters we calculate eigenfrequencies and eigenvectors of the  A$_g$- and B$_{2g}$-symmetry Raman modes in an orthorhombic YTiO$_3$ crystal and make an assignment with the phonon modes observed in the experimental Raman scattering spectra.

\section{Results and Discussion}\label{3}
\begin{figure}[b]\vspace{-0.8em}
        \includegraphics[width=120mm]{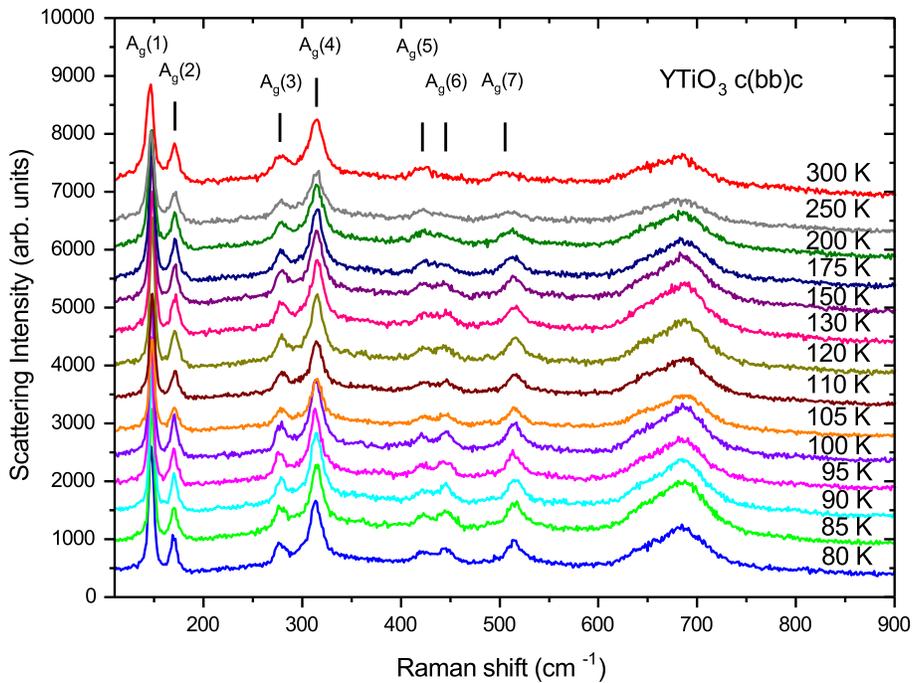}\vspace{-0.3em}
\caption{Temperature dependence (80--300\,K) of the Raman spectra measured on the oriented $ab$ surface of the YTiO$_3$ single crystal with the incident laser light polarized along the $b$ axis. The Raman spectra corresponding to sequential temperature scans are shifted along the vertical axis.}
\vspace{-0.6cm}
\label{fig1}
\end{figure}

\begin{figure}[b]\vspace{-0.8em}
        \includegraphics[width=120mm]{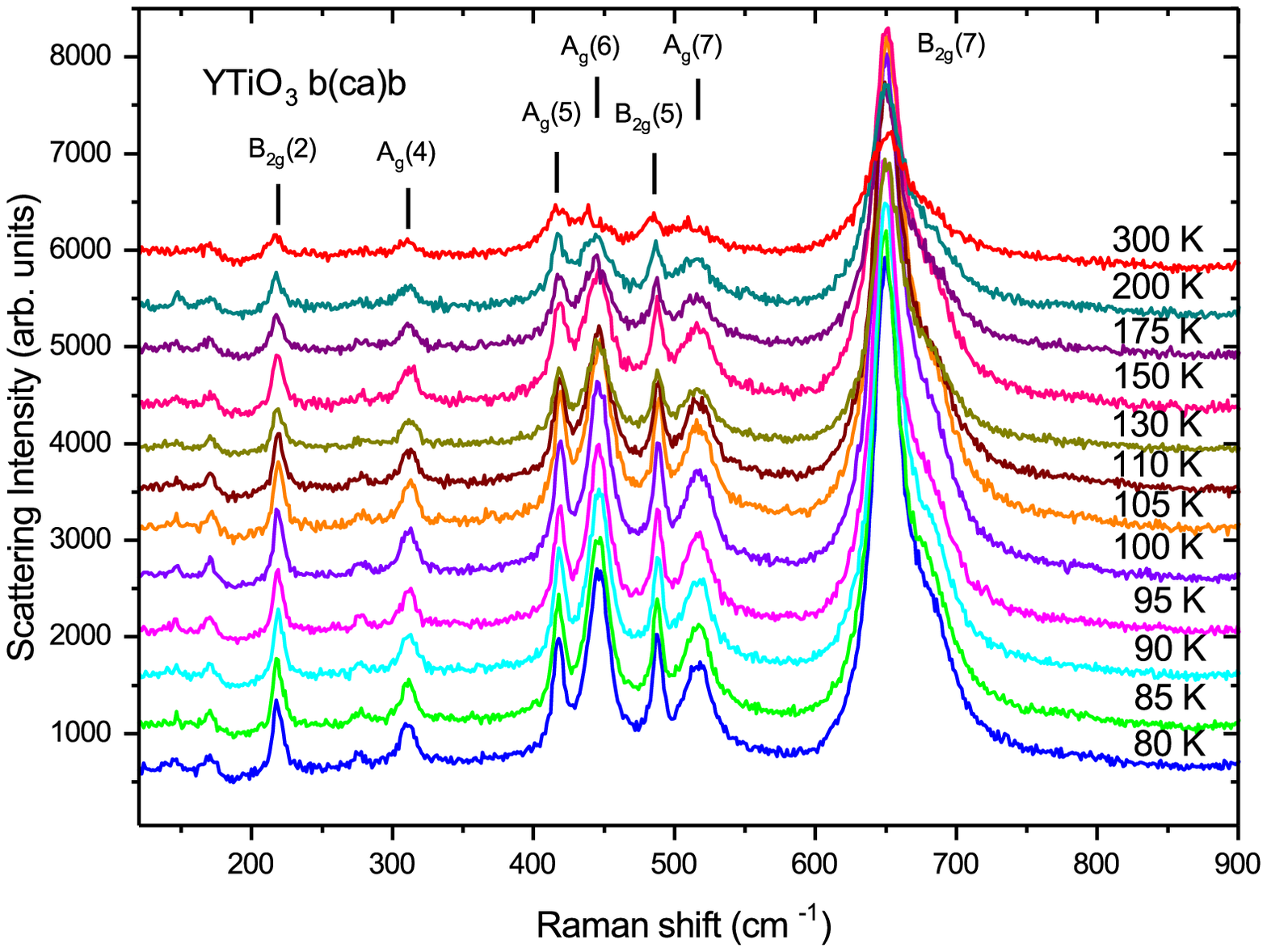}\vspace{-0.3em}
\caption{Temperature dependence (80--300\,K) of the Raman spectra measured on the oriented $ac$ surface of the YTiO$_3$ single crystal with the incident laser light polarized along the $c$ axis. The Raman spectra corresponding to sequential temperature scans are shifted along the vertical axis.}
\vspace{-0.6cm}
\label{fig2}
\end{figure}
\begin{figure}[b]\vspace{-0.8em}
        \includegraphics[width=120mm]{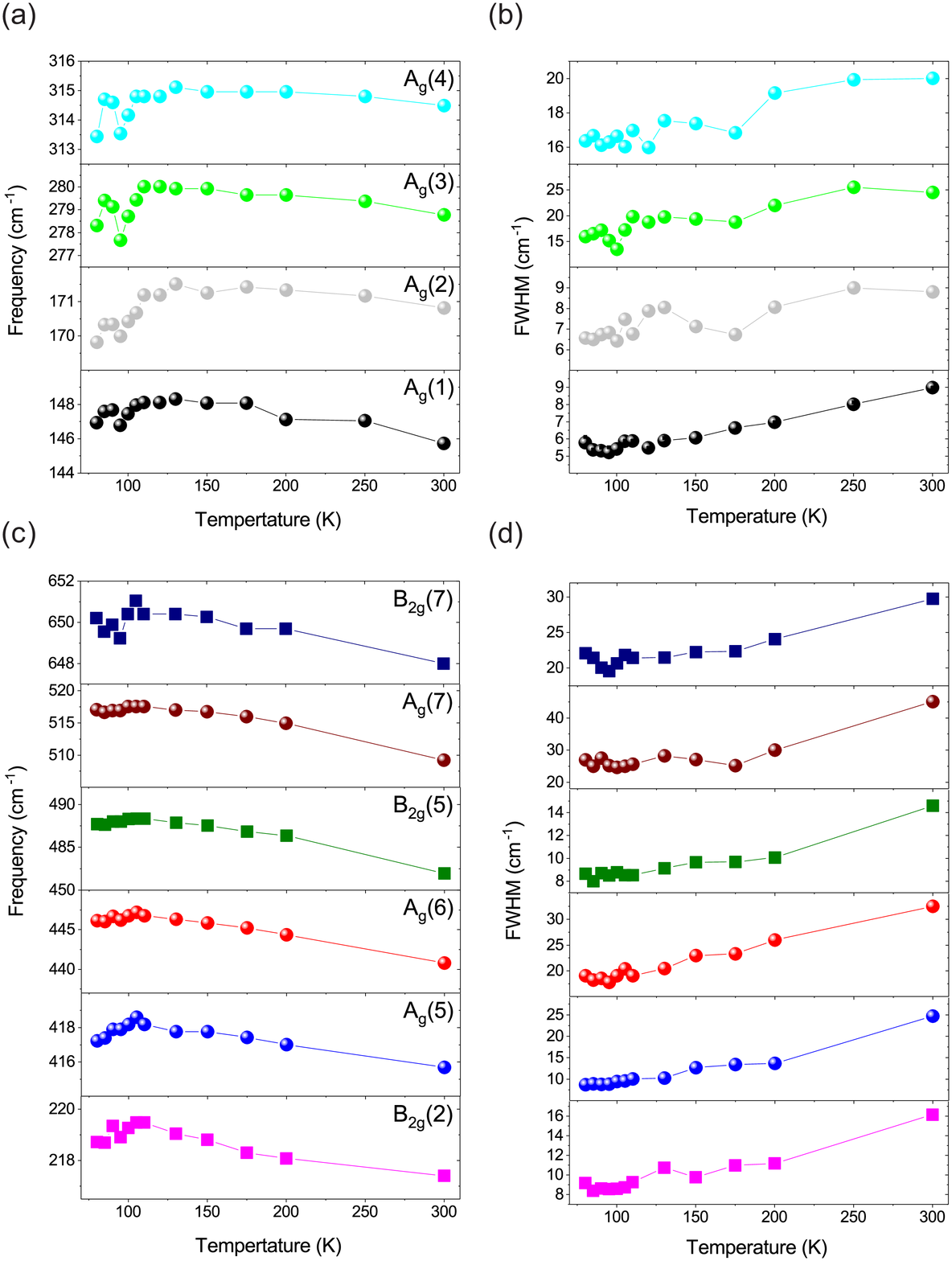}\vspace{-0.3em}
          \caption{Temperature dependences of the 
          (a,c) resonant frequencies and (b,d) FWHMs of
          some of A$_g$ and B$_{2g}$ modes resulting from the fit 
          of the Raman spectra (see Fig.\,\ref{fig1} and \ref{fig2}) 
          with Lorentzian bands.}
\label{fig3}
\end{figure}
\begin{figure}[b]\vspace{-0.8em}
        \includegraphics[width=80mm]{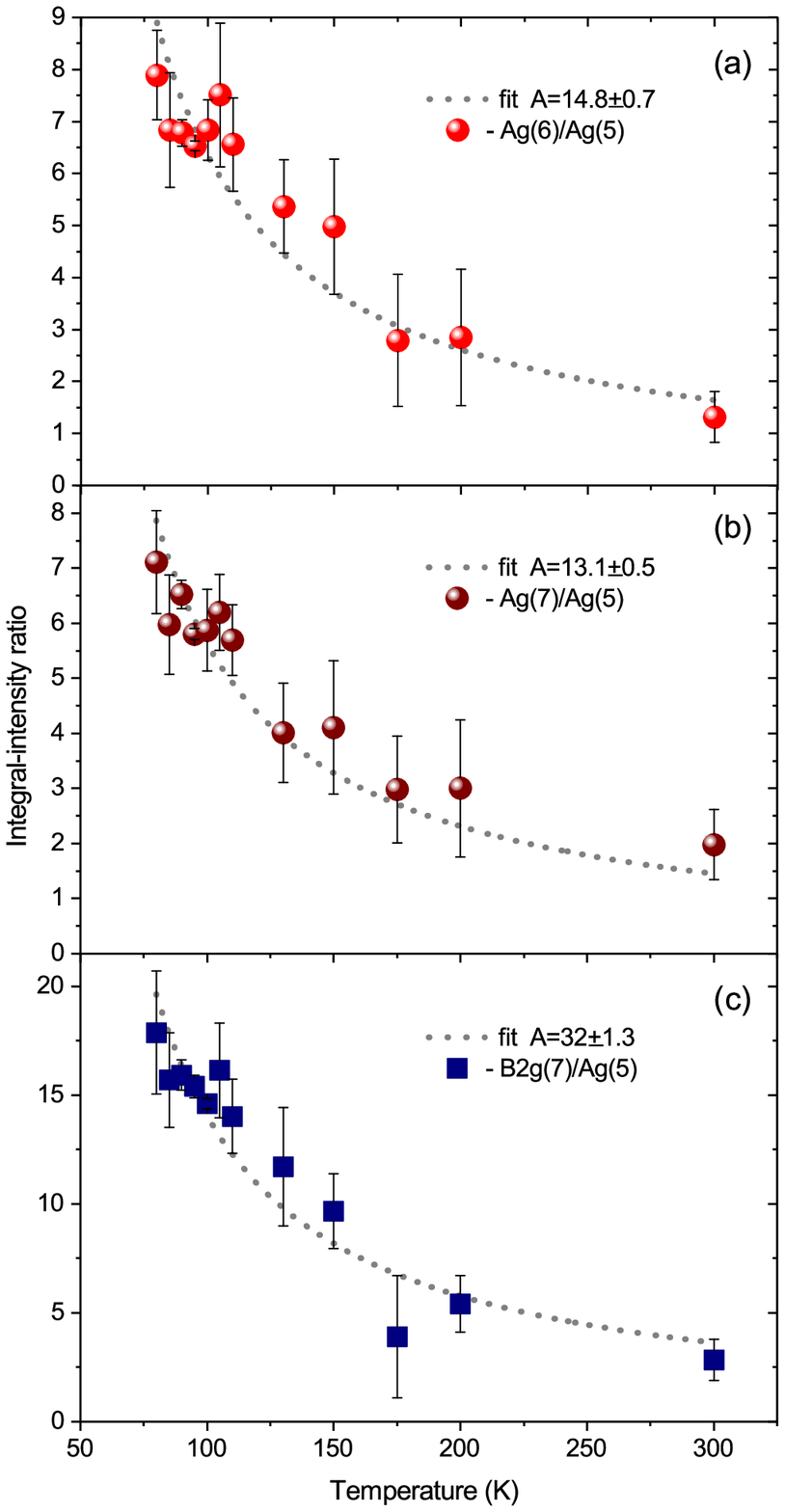}\vspace{-0.3em}
          \caption{(a-c) Temperature dependences of 
          the integral intensities ratio for the higher-frequency Raman 
          modes, fitted with the function $A\left\vert {1-\frac{T}{T_C}} \right\vert^{-1}$, where $T_C$\,=\,30 K is a critical temperature of the FM transition in YTiO$_3$.}
\label{fig4}
\end{figure}
The total number of Raman-active normal modes for the orthorhombic crystal structure of YTiO$_3$ (space group $Pnma$, D$^{16}_{2\rm h}$) with 4 f.u./unit cell is 24 (7A$_g$+5B$_{1g}$+7B$_{2g}$+5B$_{3g}$). The Raman scattering tensor settles down whose symmetry normal modes will actually appear under irradiation with polarized light at a given scattering geometry. Figures \ref{fig1} and \ref{fig2} show polarized micro-Raman spectra measured with increasing temperature from 80\,K to room temperature on the oriented (ab) and (ac) surfaces of the grown YTiO$_3$ single crystal for the two near-normal back scattering configurations c(bb)c and b(ca)b, respectively. With the incident laser light polarized along the principal $Pnma$ symmetry b-axis direction, the A$_g$ modes appear, whereas the A$_g$ and $B_{2g}$ modes were observed in the b(ca)b scattering configuration. In the low-temperature (80\,K) polarized micro-Raman spectra measured in the c(bb)c scattering configuration we identified the A$_g$-symmetry modes, A$_g$(1)--A$_g$(7), peaking at 147, 170, 278, 313, 417, 446, and 516 cm$^{-1}$ (see Fig.\,\ref{fig1}). In addition, in the low-temperature (80\,K) polarized micro-Raman spectra measured in the b(ca)b scattering configuration we were able to identify the most pronounced B$_{2g}$-symmetry modes, B$_{2g}$(2) at 218 cm$^{-1}$, B$_{2g}$(5) at 487 cm$^{-1}$, and B$_{2g}$(7) at 650 cm$^{-1}$ (see Fig.\,\ref{fig2}). We note that the A$_g$ and B$_{2g}$ modes found in the Raman spectra of the grown YTiO$_3$ single crystal are in good agreement with those of the earlier Raman study by Sugai {\it et al.} \cite{Sugai}. In the low-temperature (5\,K) polarized Raman spectra measured in YTiO$_3$ by Sugai {\it et al.}, the A$_g$ peaks were observed at 145, 168, 273, 314, 417, 446, and 512 cm$^{-1}$ and the B$_{2g}$ peaks were observed at 142, 219, 306, 328, 487, 521, and 643 cm$^{-1}$ \cite{Sugai}.

The measured temperature-dependent polarized Raman spectra (see Figs.\,\ref{fig1} and \ref{fig2}) were fitted with a set of Lorentzian bands. Figure\,\ref{fig3}(a--d) presents the resulting temperature dependences of the Lorentz peak positions and FWHMs for the observable A$_g$ and $B_{2g}$ phonon modes. One can see that the peaks of A$_g$ and $B_{2g}$ phonon modes exhibit shifts to higher frequencies with decreasing temperature from 300\,K to about 110--130\,K, while their FWHMs show narrowing, which can be associated with freezing-out of the ordinary lattice-anharmonicity effects. With further decreasing temperature down to 80\,K, most of the resonant frequencies show softening by about 2 cm$^{-1}$ (sometimes followed by hardening) and their FWHMs display a weak kink downside. 
The observed temperature softening (and hardening) of the A$_g$- and   B$_{2g}$-symmetry phonon modes below about 100\,K is associated with the temperature 
dependence of the a, b, and c ($Pnma$) lattice parameters in YTiO$_3$ crystal (see Fig.\,3(b) in Ref.\,\cite{Li}). However, to see clearly these 
trends, additional Raman measurements at lower temperatures are 
required.

The obvious effect can be recognized already from Fig.\,\ref{fig2}. There one can clearly see that the intensity of $A_g$(6):446 cm$^{-1}$ mode increases relatively to that of $A_g$(5):417 cm$^{-1}$ mode with decreasing temperature from 300\,K to 80\,K, as a result, the A$_g$(6) mode becomes predominant in the low-temperature Raman scattering spectra. Figure\,\ref{fig4}(a) 
shows the detailed temperature dependence of the integral-intensity ratio of the corresponding phonon bands, A$_g$(6)/A$_g$(5), which greatly 
increases with decreasing temperature.  
The integral intensities of $A_g$(7):517 cm$^{-1}$ and B$_{2g}$: 650 cm$^{-1}$ modes to that of $A_g$(5):417 cm$^{-1}$ mode, A$_g$(7)/A$_g$(5) and B$_{2g}$(7)/A$_g$(5), exhibit the same trend with decreasing temperature 
(see Fig.\,\ref{fig4}(b,c)). 
Surprisingly, the pronounced temperature dependences of the 
relative integral intensities develop well above the FM transition 
temperature in YTiO$_3$ ($T_C$\,=30\,K). Our analysis presented 
in Fig.\,\ref{fig4} indicates that the discovered trend may be well 
consistent with the linear behavior of the inverse susceptibility, 
1/$\chi_{mol}$ \cite{KovalevaYTO}. The origin of the observed anomalous behavior needs to be understood.
\begin{figure}[b]\vspace{-0.8em}
        \includegraphics[width=120mm]{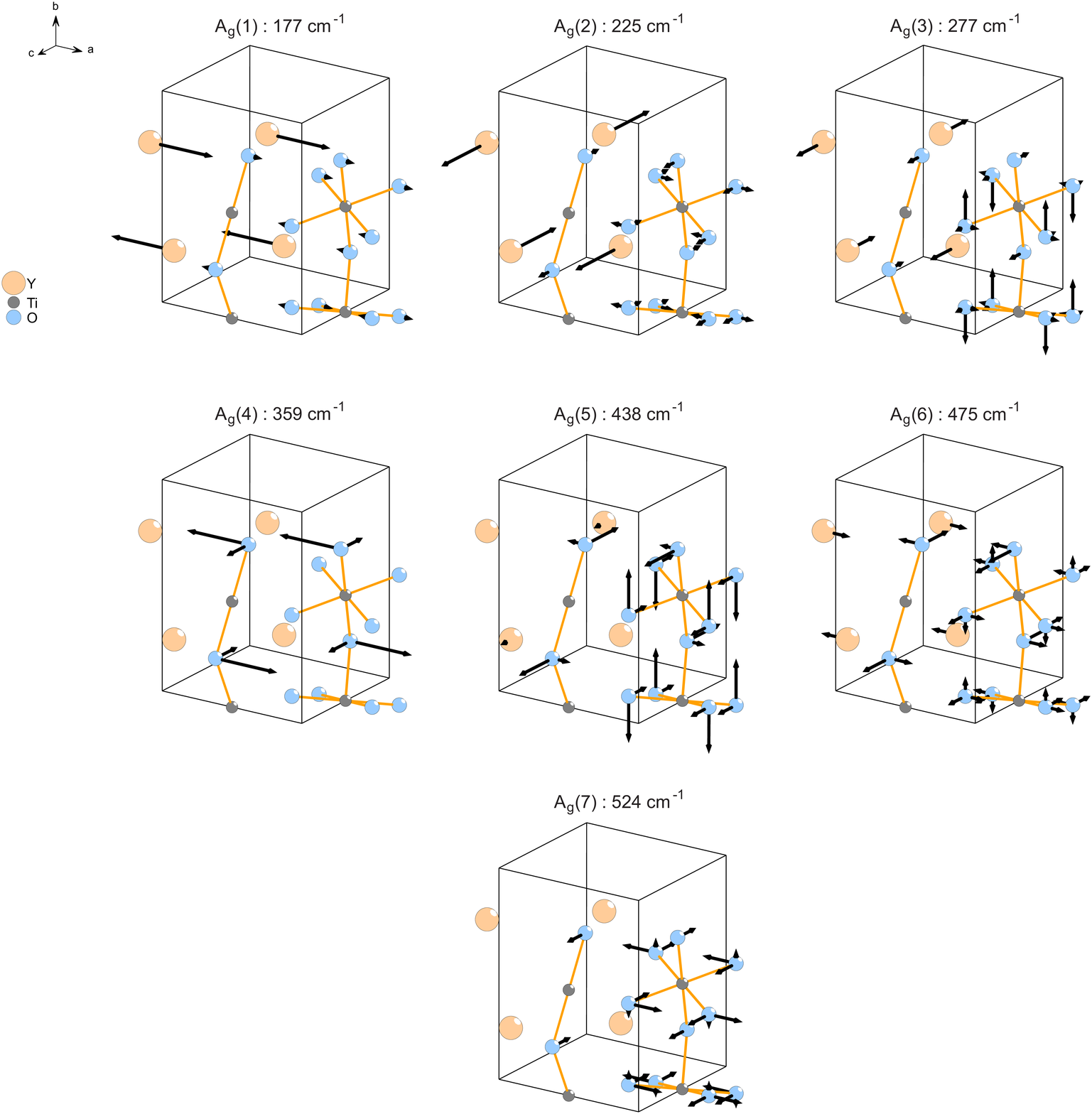}\vspace{-0.3em}
          \caption{Calculated eigenfrequencies and 
          eigenvector components of the $A_g$ symmetry modes in the orthorhombic ($Pnma$) YTiO$_3$ crystal.}
\label{fig5}
\end{figure}

\begin{figure}[b]\vspace{-0.8em}
        \includegraphics[width=120mm]{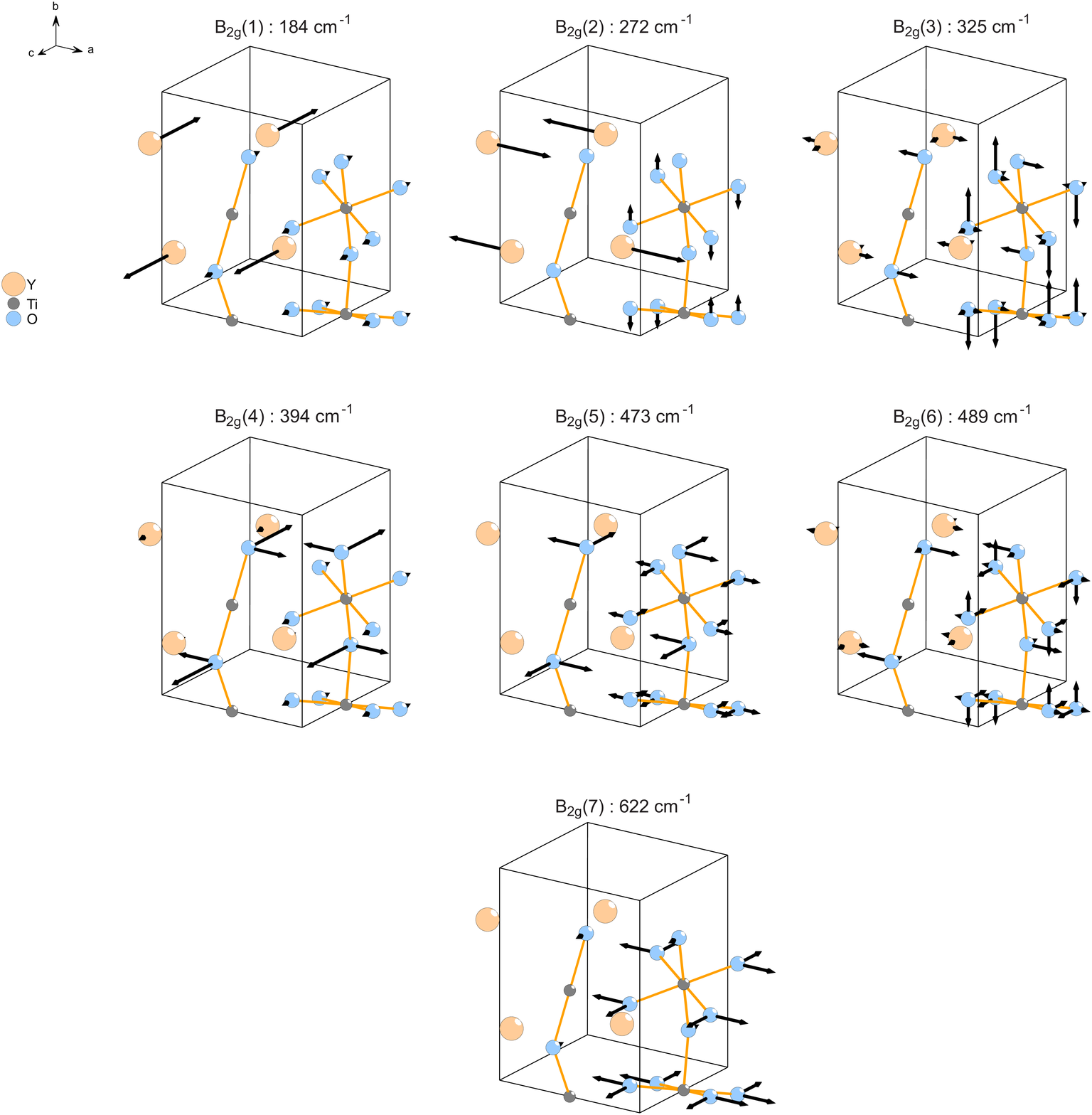}\vspace{-0.3em}
          \caption{Calculated eigenfrequencies and 
          eigenvector components of the $B_{2g}$ symmetry modes in 
          the orthorhombic ($Pnma$) YTiO$_3$ crystal.}
\label{fig6}
\end{figure}
\begin{table}
\caption{The experimental and calculated frequencies (in cm$^{-1}$) of $A_g$ and $B_{2g}$ Raman modes in orthorhombic ($Pnma$) YTiO$_3$ crystal. 
The dominant character of related atomic displacement for 
the assigned Raman modes is briefly described (more details can be seen 
from Figs.\,\ref{fig5} and \ref{fig6}).}\label{tbl1}
\begin{tabular}{llccl}
\hline
Raman mode  &\hspace{1cm}Exp.  & Exp.\,\cite{Sugai} & Calc.  &\hspace{1.5cm}Related atomic displacements \\ 
\hline
$A_g$(1) &\hspace{1cm}147&145 & 177 &\hspace{1.5cm}Y: $a$-axis\,$\parallel$ (parallel)\\
$A_g$(2)      &\hspace{1cm}170&168 & 225 &\hspace{1.5cm}Y: $c$-axis anti-$\parallel$\\
$A_g$(3)      &\hspace{1cm}278&273 & 277 &\hspace{1.5cm}Y: weak $c$-axis anti-$\parallel$\\ &&&&\hspace{1.5cm}$T_{1g}$-type ``rotational'' $Q_x$,\,$Q_y$\\
$A_g$(4)      &\hspace{1cm}313&314 & 359 &\hspace{1.5cm}O1: $a$-axis\,$\parallel$, $c$-axis anti-$\parallel$\\
$A_g$(5)      &\hspace{1cm}417&417 & 438 &\hspace{1.5cm}O1O2: $T_{2g}$-type ``scissors'' $Q_{\xi}$,\,$Q_{\eta}$\\
      &&&&\hspace{1.5cm}$T_{1g}$-type ``rotational'' $Q_x$,\,$Q_y$\\
$A_g$(6)      &\hspace{1cm}446&446 & 475 &\hspace{1.5cm}Y: weak $a$-axis\,$\parallel$\\
       &&&&\hspace{1.5cm}O1O2: weak $T_{2g}$-type ``scissors'' $Q_{\xi}$,\,$Q_{\eta}$\\
      &&&&\hspace{1.5cm}O1O2: weak $T_{1g}$-type ``rotational'' $Q_x$,\,$Q_y$\\
       &&&&\hspace{1.5cm}O2: weak in-phase stretching\\ 
$A_g$(7)      &\hspace{1cm}516&512 & 524 &\hspace{1.5cm}O2: JT $E_g$-type $Q_{\varepsilon}$\\
$B_{2g}$(1) &\hspace{1cm}&142   & 184 &\hspace{1.5cm}Y: $c$-axis\,$\parallel$\\
$B_{2g}$(2)      &\hspace{1cm}218&219 & 272 &\hspace{1.5cm}Y: $a$-axis anti-$\parallel$\\
&&&&\hspace{1.5cm}O1,O2: weak $T_{1g}$-type ``rotational'' $Q_x$,\,$Q_y$\\
$B_{2g}$(3)      &\hspace{1cm}&306      & 325 &\hspace{1.5cm}Y: weak $a$-axis anti-$\parallel$, $c$-axis\,$\parallel$\\
&&&&\hspace{1.5cm}O1O2: $T_{1g}$-type ``rotational'' $Q_x$,\,$Q_y$\\ 
$B_{2g}$(4)      &\hspace{1cm}&328      & 394 &\hspace{1.5cm}O1: $c$-axis\,$\parallel$, $a$-axis anti-$\parallel$\\
$B_{2g}$(5)      &\hspace{1cm}487&487 & 473 &\hspace{1.5cm}O1: $a$-axis\,$\parallel$, $c$-axis anti-$\parallel$\\
&&&&\hspace{1.5cm}O2: $T_{2g}$-type ``scissors'' $Q_{\zeta}$\\
$B_{2g}$(6)      &\hspace{1cm}&521      & 489 &\hspace{1.5cm}O1O2: $T_{2g}$-type ``scissors'' $Q_{\xi}$,\,$Q_{\eta}$\\
$B_{2g}$(7)      &\hspace{1cm}650&647 & 622 &\hspace{1.5cm}O2: in-phase stretching\\ \hline
\end{tabular}
\end{table}

Aimed at this purpose, here we present the results of our lattice-dynamics SM calculations and make an assignment of the Raman-active phonon modes observed in the polarized Raman spectra of YTiO$_3$ single crystal (see Figs.\,\ref{fig1} and \ref{fig2}). The eigenvector patterns for the normal modes of $A_g$ and $B_{2g}$ symmetry, as derived from the present SM lattice-dynamics calculations, are shown in Figs.\,\ref{fig5} and \ref{fig6}. Here, only major components of the eigenvectors along the principal $Pnma$ crystallographic directions are drawn for 20 atoms in the unit cell of different symmetries in all inequivalent positions. The calculated eigenfrequencies (in cm$^{-1}$) are indicated for each mode. The experimental Raman frequencies (80\,K) are assigned to the calculated frequencies of the $A_g$- and $B_{2g}$-symmetry modes in Table\,\ref{tbl1}, where the predominant character of related atomic displacements for the assigned Raman modes is briefly described according to the vibrational patterns presented in Figs.\,\ref{fig5} and \ref{fig6}. The character of the Raman modes is assigned in accordance with the prevailing contribution from the collective vibrations of Y, Ti, O1 (apical oxygen), and O2 (in-plane oxygen) atoms. One can notice from Table\,\ref{tbl1} that the calculated eigenfrequencies of the $A_g$ and $B_{2g}$ normal modes rather well agree with the experimental Raman frequencies obtained in the study by Sugai {\it et al.} \cite{Sugai}, as well as in the present Raman measurements.

For the Ti$^{3+}$ ion in $R$TiO$_3$ titanates, the ground state is described by $^2$T$_2$ term. In the $T$ problem, there are five JT-active 
modes: two tetragonal ($E$ type) modes, $Q_{\theta}$ and $Q_{\varepsilon}$, which give tetragonal distortions of cubic or tetrahedral complexes,  and three trigonal ($T_2$ type) ``scissors'' modes, $Q_{\xi}$, $Q_{\eta}$, and $Q_{\zeta}$, which give trigonal distortions of cubic complexes (these distortions of an octahedron are illustrated, for example, in Fig.\,3 of Ref.\,\cite{Mozhegorov}). The problem of the electronic $T$-states coupled by $E$ and $T$ modes was discussed earlier \cite{OpikPryce}. Let the strength of the linear electron-lattice coupling in the $T$ problem is measured by coupling coefficients $G_{TE}$ and $G_{TT}$, and the corresponding force constants are $K_E$ and $K_T$, respectively. It was shown that the tetragonal ($E$ mode) distortions will give the minimum energy configuration if $2G^2_{TE}$/$K_E>\frac{2}{3}G^2_{TT}/K_T$, otherwise the trigonal ($T$ mode) distortions will give the minimum energy configuration. Therefore, usually the minimum energy configuration involves only one of the modes, while another one gives a saddle point in the potential energy surface (unless there is an accidental coincidence, $2G^2_{TE}$/$K_E=\frac{2}{3}G^2_{TT}/K_T$, and the minimum energy configuration is balanced by some subtle conditions). The $T_{2g}$-type ``scissors'' $Q_{\xi}$ and $Q_{\eta}$ distortions are strongly pronounced in the $A_g$ (5) mode at 417 cm$^{-1}$, and $Q_{\zeta}$ distortion is strongly pronounced in the $B_{2g}$(5) mode at 487 cm$^{-1}$ 
(see Figs.\,\ref{fig5} and \ref{fig6} and Table\,\ref{tbl1}). However, we observe that the $A_g$(5) mode associated with the ``scissors'' $Q_{\xi}$ and $Q_{\eta}$ distortions becomes significantly suppressed with decreasing temperature, especially below about 100\,K (see Figs.\,\ref{fig2} and \ref{fig4}(a)). Instead, its higher-frequency satellite, $A_g$(6) mode peaking at 446 cm$^{-1}$, notably increases in intensity. According to our assignment, this mode shows a mixed character with many weak contributions, including the ``rotational'' ($T_{1g}$-type), $Q_x$, $Q_y$, $Q_z$, trigonal ($T_{2g}$-type), and tetragonal ($E_g$-type) distortions (for details, see Fig.\,\ref{fig5} and Table\,\ref{tbl1}). In addition, the $A_g$(7) mode at 517 cm$^{-1}$ associated with the JT $E_g$-type tetragonal $Q_{\varepsilon}$ distortions notably increases in the intensity. The same observation is pertinent for the strong $B_{2g}$(7) mode at 650 cm$^{-1}$, which is assigned to the O2 (in-plane) oxygen in-phase stretching (see Figs.\,\ref{fig5} and \ref{fig6} and Table\,\ref{tbl1}). This leads us to the conclusion that the ground state resulting from the trigonal distortions becomes unstable against the tetragonal distortions. It is supposed the GdFeO$_3$-type distortion in $R$TiO$_3$ lifts the $t_{2g}$ orbital degeneracy into three non-degenerate levels, and the lowest orbital occupation stabilizes the G-type AFM state \cite{Mochizuki}. The two higher nearly-degenerate states may still possess the JT instability. In the $3d(t^1_{2g})$ series of orthorhombic $R$TiO$_3$ titanates, the JT instability is the most pronounced in YTiO$_3$ compound. We believe that the obtained in the present study evidence on the relative enhancement of the contributions of the $E_g$-type to $T$-type modes indicates that the JT instability existing between these two nearly-degenerate electronic levels depletes the $Y$-ion $D_{3d}$  crystal field (GdFeO$_3$-type distortion) effects in YTiO$_3$. 
However, the temperature behavior near the magnetically ordered state will be largely determined by the interplay between the spin-orbit coupling and the JT effect. The resulting rearrangement and/or refinement of the orbital structure indicates the route towards the establishment of the FM phase transition at $T_C$\,=\,30\,K. 

\section{Conclusion}\label{4}
The observed anomalous behavior of the resonant frequencies and widths of the $A_g$- and $B_{2g}$- symmetry Raman modes well correlates with significant deviation of the inverse susceptibility, 1/$\chi_{mol}$, from the Curie-Weiss mean-field behavior found in YTiO$_3$ crystal below 100\,K \cite{KovalevaYTO}. The vibrational character of the higher-frequency Raman modes due to ``rotational'' ($T_{1g}$-type), trigonal ($T_{2g}$-type), and tetragonal ($E_g$-type) distortions was clarified from the theoretical shell-model calculations. 
The enhancement of the $A_g$-mode assigned to the Jahn-Teller (JT) tetragonal $E_g$-type $Q_{\varepsilon}$ distortion with respect to the trigonal $T_{2g}$-type ``scissors'' $Q_{\xi}$ and $Q_{\eta}$ distortions led us to the conclusion that trigonal GdFeO$_3$-type distortions become unstable against tetragonal Jahn-Teller distortions with decreasing temperature from 300\,K down to about 100\,K. The following observed phonon anomalies indicate the orbital 
structure refinement on the route to the FM phase transition in YTiO$_3$ at $T_C$\,=\,30\,K.\\ 

{\bf Acknowledgement}\\

The authors thank J. Gale for making actual General Utility 
Lattice Program (GULP) used in the present shell-model 
calculations. The authors acknowledge Prof. B. Keimer and Max-Planck Society for the possibility to perform our Raman measurements on the YTiO$_3$ 
single crystal grown at the Max-Planck Institute 
for Solid State Research in Stuttgart.\\  
 
{\bf Declaration of competing interest}\\

The authors declare that they have no known competing financial interests or personal relationships that could have appeared to influence the work reported in this paper.

\end{document}